\documentclass[12pt,twoside]{article}  
\usepackage{fleqn,espcrc1}

\newcommand{\be}{\begin{equation}}  
\newcommand{\ee}{\end{equation}}
\newcommand{\bea}{\begin{eqnarray}}
\newcommand{\eea}{\end{eqnarray}}
\newcommand{\bitem}{\begin{itemize}\setlength{\itemsep -5pt}}
\newcommand{\eitem}{\end{itemize}}
\newcommand{\ombe}{\Omega_B}
\newcommand{\omnu}{\Omega_{\nu}}
\newcommand{\omst}{\Omega_{\star}}
\newcommand{\omm}{\Omega_m}
\newcommand{\omlam}{\Omega_{\Lambda}}
\newcommand{\omx}{\Omega_X}
\newcommand{\bc}{\begin{center}}
\newcommand{\ec}{\end{center}}

\newcommand{\AmS}{{\protect\the\textfont2
  A\kern-.1667em\lower.5ex\hbox{M}\kern-.125emS}}
\title{VACUUM ENERGY: IF NOT NOW, THEN WHEN?}

\author{Sidney A. Bludman\address{DESY, 22607 Hamburg, Germany \\
        University of Pennsylvania, Philadelphia, PA 19104}
       \thanks{Supported in part by Department of Energy grant DE-AC02-76-ERO-3               071.  \quad \quad \quad \quad \quad {\bf DESY 99-089}}}

\begin{document} 
\maketitle  
\begin{abstract}
We review the cosmological evidence for a low matter density universe
and a cosmological constant or dynamical vacuum energy and address the cosmological 
coincidence problem: why is the matter density  about one-half the vacuum
energy {\em now}. This is reasonble, following the  anthropic argument of Efstathiou 
and of Martel, Schapiro \& Weinberg. 
\end{abstract}

\section{DENSITY OF RADIATION AND CLUSTERED MATTER IS LOW [1,2]} 

\subsection{Different Kinds of Dark Matter and Vacuum Energy}

Radiation, massive neutrinos, luminous matter 
are now known to consitute the least part of the Universe:
$\Omega_{rad}=5\times10^{-5}h^{-2},~\omnu=(0.003-0.15) \sim \omst=(0.005 \pm
0.002)$.  Instead, we find 
most of the Universe in three forms: 
\bitem
\item Baryonic Matter: $\ombe=(0.019\pm 
0.0012)h^{-2} \sim (0.05\pm 0.005)   >>\omst$
\item Non-baryonic Cold 
Dark Matter: $\omm=(0.35 \pm 0.1)h^{-1/2} \sim (0.4 \pm 0.1)
>>\ombe$
\item Smooth Energy: $\omx \sim 0.6 \pm 0.2$. 
\eitem
Here the various $\Omega_i$ 
are the fractional contributions to the present critical density $\rho_{cr} \equiv 
3H_0^2/8 \pi G$.
 
From the cosmic background radiation (CBR), we know that
the spatial curvature $\Omega_k \equiv 1-\omm - \omx \pm 0.2$, a flat 
universe of critical density, dominated by non-baryonic 
matter and smooth energy. Inflation explains this flatness of the present 
universe, as a consequence of an early de Sitter phase making an old 
universe, now expanding slowly. 

We distinguish two kinds of dark matter relicts of the Big Bang:
(1) Hot Dark Matter (HDM), light neutrinos $\sum m_{\nu} < 45~eV$.  Because
we know large-scale structure (LSS) evolved hierarchically,  $\omnu<0.15$, 
HDM cannot dominate cosmologically; (2)Instead, Cold Dark Matter (CDM),
say $10^{-5 \pm 1}~eV$ axions or  $(50-500)~GeV$ neutralinos must seed
the gravitational clustering that we see.  Indeed, 2/3 of our Galactic halo 
DM cannot be baryonic, and must be CDM.

\subsection {The Density of Clustered Matter: $\omm \sim 0.4$} 

The observed primordial abundances of the four light nulei are concordant
with a single baryon density \be \Omega_B =(0.019 \pm 0.0012)h^{-2}
=0.05 \pm 0.01 . \ee  This value is consistent with $\Omega_B h^2 > 0.015$
from large scale structure (LSS) and with the height of the 
first acoustic peak in cosmic background radiation angular distribution.

In rich clusters of galaxies, most baryons reside in the hot inter-galactic
gas and and reveal themselves in inner bremsstrahlung X-rays and in Sunyaev-
Zeldovich infra-red radiation. These two sources determine the total baryon
fraction to be  $f_B=(0.075 \pm 0.002)h^{-3/2}$ and $(0.06 \pm 0.006) 
h^{-1}$ respectively.   Together with the above value of $\Omega_B$, this
determines the clustered mass density \cite{Blud98}\be \omm \equiv \ombe/f_B=
(0.4 \pm 0.1). \ee

The simplest argument for mean background mass density $\omm > 0.3 (95\% 
C.L.)$ derives from limits on the outflow from cosmic voids.
Supporting evidence also come from
\bitem
\item The evolution of the number density of rich clusters $\omm=(0.45 \pm 0.1) $;
\item Peculiar velocities of IRAS galaxies;
\item Large Scale Structure, which imprints the present mass density 
power spectrum, showing $ \omm h=(0.25 \pm 0.05)$ ;
\item $Ly-\alpha$ absorption spectra compared with LSS simulations when growth
was becoming non-linear, showing     $\omm = 0.34 \pm 0.1$ .
\eitem
 
\section{ACCELERATED EXPANSION REQUIRES VACUUM ENERGY} 

Different kinds of energy density with equation of state $w \equiv P/\rho$,
expand adiabatically with scale $a$ at different rates $\rho \sim a^{-n},~n \equiv
3(1+w)$:\\
\begin{center}
\begin{tabular}{|l|c|c|}  \hline
{\em substance}          &  {\em w}  &  {\em n}  \\  \hline

radiation &  1/3   &  4   \\
NR matter &   0    &  3  \\
curvature & -1/3   &  2  \\
vacuum    & -1     &  0   \\
\hline  
\end{tabular}  \\
\end{center}
In this list, we have also included the spatial curvature term
which falls off as $a^{-2}$, just as a substance with 
$\rho+3P=0$.

\subsection{Expansion Age Requires Vacuum Energy}  

The oldest globular clusters are aged $11.5 \pm 1.3~Gyr$.   Allowing
about $1~Gyr$ before star formation at $z \sim 5$, this gives a dynamical 
age $t_0 \geq 14~Gyr$.  Because the Hubble constant is now known
$H_0=73 \pm 6(stat) \pm 8(syst)$, $H_0 t_0 \geq 0.93 \pm 0.16$.
The $1\sigma$ limits rule out an Einstein-deSitter $\omm=1,\omlam=0$ 
flat universe, but are consistent with a low matter density $\Omega_m 
\sim 0.4$ and a cosmological constant $0.3< \omlam < 1.3$.
The two extreme $\omm=0.4$ geometries, open ($\omlam=0,~H_0 t_0=0.80$) and 
flat ($\Omega_\Lambda=0.6,~H_0 t_0=0.89$), are
distinguishable by different cosmographic distance measurements:
\bitem
\item A difference in lumininosity distance, observable with SN Ia;
\item Different comoving volume elements, observable in the 
number density of gravitational lenses;  
\item Different differential intersection probabilities, observable in QSO 
absorption lines or pencil-beam redshift surveys.
\eitem
We now want to fix the present values of the two cosmological
parameters, $\Omega_\Lambda$ and $\Omega_m$ separately. 

\subsection{The Young Universe Was Expanding {\em Slower} Than Now}

A selected sample of supernovae Ia (``Branch normal'') are ``standard
bombs'', from which the luminosity distance $d_L$ can be measured at
high red-shift $z>0.4$. From more than 40 such SN Ia, the combination
$\Omega_\Lambda-4 \Omega_m/3 =1/3 \pm 1/6 $ is measured [4,5].  The
nearly orthogonal combination $\omm + \Omega_\Lambda \sim 1, $ gives
the joint fit $\omm \sim 1/3, \quad \Omega_\Lambda \sim 2/3.$ More
precisely, what is established is that the present deceleration $
q_0=(\omm + (1+3w_X)\omx)/2 <0 $, because of a component $X$, with
$w_X<-1/2.$ This shows that a positive pressure $w>0$ would not cause,
but retard, cosmic expansion.  Any exotic component with $w_X<-1/3$
violates the once-respected energy positivity condition $\rho+3P<0$,
and accelerates expansion, acting as a negative (accelerating) active
gravitational mass.  Weaker supporting evidence comes from double-lobe
radio galaxies, from rich clusters, and from the frequency of
multiply-lensed QSOs.

Because the supernovae observations are so important, we must remember
that what is observed is only that high redshift supernova are
fainter by 0.25 magnitude than they would appear in an accelerating
universe. Possible systematic errors that might produce the same
apparent (13\%) dimming are \bitem
\item Part of the dimming may be caused by intervening (gray!) dust;
\item Nearby supernovae may have brightened with age.  We now need
more {\em nearby} standard Ia 's for calibration.
\eitem

\section{QUINTESSENCE: DYNAMIC VACUUM ENERGY} 

The present supernovae data does not require a cosmological constant
$\rho_\Lambda=-P_\Lambda =$ constant, but only some unclustered {\em
  smooth dark energy} or {\em quintessence}, with $w_X<-1/2$
\cite{Wang98}. This ``time-dependent vacuum energy'', may be no more
than a scalar field $\rho=(1/2)\dot{\phi}^2
+V(\phi),~P=(1/2)\dot{\phi}^2 - V(\phi)$ slowly rolling down towards
its true ground state. If the potential energy $V(\phi)\geq $ thrice
the kinetic energy, the time averaged $ -1<w_{eff}<-1/2$, and $\rho_X$
diminishes slower than $a^{-3/2}$ and, since $z<(0.14-0.3)$, already
dominates the curvature and matter energy density.

Such a scalar field is reminiscent of the McCrae-Hoyle C-field, with
two important differences: (1) Steady state cosmology requires
continuous creation, violates local energy conservation and is ruledd
out by baryogenesis, nucleogenesis, cosmic background radiation and
the observed evolution of galaxies and QSOs; (2) Quintessence is
consistent with the conservation laws of General Relativity.

\section{CONCLUSIONS:  $\Omega_X \sim \omm $ IS 
REASONABLE FOR OUR UNIVERSE} 

Inflation never determined the matter and vacuum energy content of the
universe, other than requiring $\rho_m+\rho_\Lambda=\rho_{cr}$ .
There (fortunately) never was any theory predicting $\Lambda=0$, nor any 
symmetry principle protecting this value:  a rolling scalar
field or any value $\omlam<1$ is reasonable.  The real cosmological problem 
is not why $\rho_\Lambda$ or  $\rho_X<1$ are present, but to explain
the cosmic cooincidence, that  
the matter density $\rho_m$, which is ever-diminishing has the value
$\sim \omlam/2$ {\em now}. 

While we hope for a future theory of everything that will calculate all 
the constants of particle physics, $\omlam $,
alone of all constants, may not be calculable  from first principles.  Instead, we
ask what values for $\Omega_{\Lambda}/\Omega_m$ would be reasonable 
{\em cosmologically}.

\subsection{``If not now, then when?'' [6]}

Efstathiou \cite{Efs95} and Martel, Shapiro \& Weinberg \cite{Mar98} 
answered this question, before the supernovae observations establishing a constant or time-varying
vacuum energy. If $\omlam $ is a stochastic variable,
the probability distribution of its observed values should be proportional to
the fraction of matter 
destined to condense into galaxies, stars, astronomers.  They find that
the probability distribution for random living astronomers observing a value
$\omlam $, in a nearly flat universe, is broadly distributed about a median $\omlam  
\sim 0.75$:
(1) If $\omlam /\omm >4$, the universe would have expanded
too fast to form such structures; (2)  $\omlam /\omm <0.02$ is possible, but
relatively unlikely; (3) Values $\omlam /\omm \sim 2$ would be
observed 5-12\% of the time. The observed value,
while below the median in probability, is close to the maximum value consistentwith the formation of galaxies by the present epoch.
\subsection{The Future}

We live at the only time we could be living, at the end of the epoch of 
galaxy and star formation.   Barring
another (unlikely) particle physics phase transition, the universe is moving
monotonically toward a de Sitter universe fixed point. 

\subsection{The Nature of Scalar Fields}

Higgs scalars and 
vacuum expectation values and masses enter mysteriously 
into present-day broken symmetry gauge theories. 
A rolling scalar
field with very low mass  ($<10^{-3}~eV$) is an important mediator of the
longest range forces, which may be telling us something: the 
anthropic principle recognized in cosmology, may become the heuristic
guide particle physics needs.

\end{document}